\documentclass[12pt,a4paper]{article}
\usepackage{epsfig} 
\usepackage{graphics}
\usepackage{subfigure}
\usepackage{a4wide}
\usepackage{amsfonts}

\title{Hopf Solitons on the Lattice}
 \author{R S Ward\footnote{email: richard.ward@durham.ac.uk}
 \bigskip
\\Department of Mathematical Sciences,  \\ University of
Durham, \\Durham DH1 3LE}

\newcommand{\RR}{{\mathbb{R}}}
\newcommand{\ZZ}{{\mathbb{Z}}}
\newcommand{\pa}{\partial}

\begin{document}

\maketitle \abstract{Hopf solitons in the Skyrme-Faddeev model ---
$S^2$-valued fields on $\RR^3$ with Skyrme dynamics --- are string-like
topological solitons. In this Letter, we investigate the analogous lattice
objects, for $S^2$-valued fields on the cubic lattice $\ZZ^3$ with a
nearest-neighbour interaction. For suitable choices of the interaction,
topological solitons exist on the lattice.  Their appearance is remarkably
similar to that of their continuum counterparts, and they exhibit the same
power-law relation $E\approx c H^{3/4}$ between the energy $E$ and the
Hopf number $H$.
}


\newpage


One of the simplest 3-dimensional systems admitting topological solitons is
the O(3) sigma model --- in other words, where the field is a unit 3-vector
$\vec{\psi}(x^j)$ on $\RR^3$.  Given the usual boundary condition
$\vec{\psi}(x^j)\to\vec{\Psi}$ (constant) as $|x^j|\to\infty$, such configurations
are classified by their Hopf number $H[\psi]\in \pi_3(S^2)\cong\ZZ$.
There is an integral formula for $H[\psi]$: first define
$F_{jk} = \vec{\psi}\cdot(\pa_j\vec{\psi})\times(\pa_k\vec{\psi})$;
then find $A_j$ such that $F_{jk}=\pa_j A_k - \pa_k A_j$ (this is always
possible); and finally, compute
\begin{equation} \label{charge}
  H[\psi] = \frac{1}{32\pi^2} \int_{\RR^3} \varepsilon^{jkl} F_{jk} A_l \,d^3x.
\end{equation}

An energy functional of the form $E=\int\bigl[|\pa\psi|^2+V(\psi)]\,d^3x$
does not admit stationary soliton solutions with nonzero $H[\psi]$:
for example, $E$ has no smooth critical points, and there are no stationary
solutions of the Landau-Lifshitz equation \cite{P93}.  But with a modified
$E$, one can have interesting soliton solutions.  The best-known example
comes from adding a Skyrme term \cite{F75, FN97}, so that
\begin{equation} \label{energy}
 E[\psi]=\int \bigl[ (\pa_j\vec{\psi})\cdot(\pa_j\vec{\psi})
         + \frac{1}{4} F_{jk} F_{jk} \bigr] \, d^3x.
\end{equation}
It is then believed that $E$ has a smooth minimum in each topological
class, with $E\approx c H^{3/4}$ for some constant $c$; and there is
considerable evidence supporting this conjecture, both analytic
\cite{VK79, W99, LY04} and numerical \cite{GH97, BS99, HS99, HS00}.
These solitons can be
visualized as closed curves, which may be linked or knotted. In particular,
if $p$ is the point on $S^2$ corresponding to the boundary value $\vec{\Psi}$,
and $q\in S^2$ is the antipodal point, then the soliton may be viewed as the
closed curve $\psi^{-1}(q)$
in $\RR^3$, which links $|H|$ times around the open curve $\psi^{-1}(p)$.

In this Letter, we consider a lattice rather than a continuum system:
namely, the field consists of a unit vector $\vec\phi$ defined at
each point of the three-dimensional cubic lattice $\ZZ^3$. At first sight,
one then loses all the topological features; but by restricting to
``well-behaved'' configurations, many of the topological properties can be
maintained.  There are many well-known examples of this, in other systems
admitting topological solitons. One such is the
O(3) model on the 2-dimensional lattice $\ZZ^2$: for a suitable
set of configurations, one can define the topological charge
\cite{BL81}, and (depending on the choice of dynamics) one can have a
Bogomolny bound on the energy, and the existence of stable topological solitons
\cite{W95, W97}.  We shall see below that analogous results hold for the
3-dimensional case.  In particular, it makes sense to talk about
Hopf solitons on $\ZZ^3$; it turns out that they resemble their continuum
counterparts, and that their energy has the same behaviour $E\approx c H^{3/4}$
as in the continuum case.


So let us consider a spin (unit 3-vector) $\vec\phi=\vec\phi_{i,j,k}$ defined at
each site of the three-dimensional cubic lattice $\ZZ^3$. Equivalently,
one may think of a map $\phi:\ZZ^3\to S^2$.  The boundary
condition at infinity is $\vec\phi_{i,j,k}\to\vec\Phi:=(0,0,1)$ as
$i^2+j^2+k^2\to\infty$.  We assume there to be an isotropic nearest-neighbour
interaction, so the energy $E[\phi]$ of a configuration has the form
\begin{equation} \label{lattice-energy}
 E[\phi] = \sum_{i,j,k} \left[f(\vec\phi_{i+1,j,k}\cdot\vec\phi_{i,j,k})
    + f(\vec\phi_{i,j+1,k}\cdot\vec\phi_{i,j,k})
    + f(\vec\phi_{i,j,k+1}\cdot\vec\phi_{i,j,k}) \right],
\end{equation}
where $f$ is a suitable function. We may take $f(1)=0$, so that the constant
field $\vec\phi_{i,j,k}\equiv\vec\Phi$ has zero energy; and $f'(1)=-1$,
which sets the energy scale.

The simplest such function is $f(\xi)=1-\xi$, which corresponds to the
usual Heisenberg model; but for this choice of $f$, the only minimum
of $E[\phi]$ is the constant field $\phi(x^j)\equiv\Phi$. In order for interesting
local minima of the energy to exist, we need to make it energetically more
unfavourable for $\phi$ to become ``discontinuous'', in the sense that neighbouring
spins point in wildly different directions. For example, this can be achieved
by taking $f(\xi)=(1-\xi)+\alpha(1-\xi)^2$, with the parameter $\alpha$ being
sufficiently large.

Another way to view the situation is that we want to restrict to ``continuous''
configurations $\phi$, for which a Hopf number $H[\phi]\in\ZZ$ can
be unambiguously defined. This is a familiar idea: as mentioned above,
for example, one can
define the winding number of a spin-field on $\ZZ^2$ by excluding certain
exceptional configurations \cite{BL81}.  In the present case, one may think
of $\ZZ^3$ as embedded in $\RR^3$; if $\phi:\ZZ^3\to S^2$ can be extended to
a continuous map $\psi:\RR^3\to S^2$, in a way which is unambiguous up to
homotopy, then $H[\phi]$ can be defined as $H[\psi]$.  In order for such
an unambiguous interpolation $\psi$ to exist, we need a ``continuity''
condition on $\phi$. Let us impose the following condition: that the angle
between any pair of nearest-neighbour spins is acute.  Continuous fields
$\phi$ are those which satisfy this condition:
$\vec\phi_{i+1,j,k}\cdot\vec\phi_{i,j,k}>0$ for all $i,j,k$, with a similar
inequality for the $j$-links and the $k$-links.

To see what might happen if $\phi$ becomes discontinuous, consider
a face of the lattice, say $P=\{(0,0,0),(1,0,0),(0,1,0),(1,1,0)\}$.
Its image on $S^2$ is a spherical quadrilateral. If neighbouring spins
were allowed to be orthogonal, then the four vertices
$\{\vec\phi_{0,0,0}, \vec\phi_{1,0,0}, \vec\phi_{0,1,0}, \vec\phi_{1,1,0}\}$
of this quadrilateral could lie on the equator of $S^2$.  But then it would
be unclear how to interpolate: should the image of the interior of the face
$P$ be the northern hemisphere of $S^2$ or the southern hemisphere?  For
continuous fields this ambiguity cannot occur, and the interpolation
$\psi$ is well-defined up to homotopy.  In the two-dimensional case (maps
from $\ZZ^2$ to $S^2$), one can then define the winding number directly, by
adding up the signed areas ${\cal F}(P)$ of the spherical quadrilaterals
corresponding to each face \cite{BL81}.  For the three-dimensional case,
computation of $H[\phi]$ is not quite so straightforward: we shall return to
this below. The basic fact is that, just as in the continuum case, the space
of continuous configurations is disconnected, and its components are labelled
by the Hopf number $H[\phi]\in\ZZ$.

As mentioned above, the simplest choice $f(\xi)=1-\xi$ of inter-spin
potential does not permit the existence of static solutions (local minima
of $E$) with non-zero Hopf number $H$. If one starts with a continuous
configuration $\phi$ having $H[\phi]\neq0$, and allows it to flow down
the energy gradient, then $\phi$ becomes discontinuous, and the topology is
lost.  If this is to be avoided, the function $f(\xi)$ needs to contain
higher powers of $1-\xi$.  The simplest choice is
\begin{equation} \label{fxi}
   f(\xi) = (1-\xi)+\alpha(1-\xi)^2 \,.
\end{equation}
The parameter $\alpha$ has to be large enough to avoid the instability
referred to above, and numerical experiments indicate that a value of
$\alpha=25$ is sufficient for this (whereas $\alpha=16$, say, is not).
In what follows, we adopt (\ref{fxi}) with $\alpha=25$.

The system defined by (\ref{lattice-energy}, \ref{fxi}) was investigated
numerically: the procedure consisted of mimimizing the energy $E$,
using a conjugate-gradient method.
The quantity $\min\{\vec\phi\cdot\vec\phi_+\}$ (the minimum being
taken over all links) was monitored, to check that it remained
positive --- in other words, that $\phi$ remained continuous as
it flowed down the energy gradient.  The computation was done for finite
lattices of size $N^3$, with $\vec\phi$ set equal to $\vec\Phi=(0,0,1)$ at
the boundary of the lattice, for a range of values of $N$ up to $N=22$;
the results were then extrapolated to the unbounded ($N\to\infty$) limit.  

In addition, the Hopf number $H[\phi]$ was monitored.
One can get an approximate value for $H[\phi]$ by using
a discrete version of the formula (\ref{charge}), as follows. The first step
is to compute, for each face $P$ of the lattice, the signed area ${\cal F}(P)$
of the corresponding spherical quadrilateral.
This object ${\cal F}$, which assigns a number ${\cal F}(P)$ to each face $P$,
is a 2-cocycle on $\ZZ^3$, and is the analogue of the 2-form $F$ in
(\ref{charge}).  Given ${\cal F}$, one can compute a 1-cochain ${\cal A}$
satisfying $\delta{\cal A}={\cal F}$; this object ${\cal A}$ assigns a number
${\cal A}(L)$ to each link $L$, and is the analogue of the 1-form $A$ in
(\ref{charge}).  Then
\begin{equation} \label{lattice-charge}
  H_0[\phi] = \frac{1}{128\pi^2} \sum_{P\perp L} {\cal F}(P)\,{\cal A}(L)
\end{equation}
gives an approximation to $H[\phi]$.  By interpolating $\phi$ to a lattice with
half the lattice spacing ({\it ie.}\ with eight times as many sites) and
repeating the calculation, one gets a better approximation $H_1[\phi]$; and
this procedure can be iterated to obtain a sequence $H_n$ which converges to $H$.
In fact, the quantity $H'=(4H_1[\phi]-H_0[\phi])/3$ is already within
$1\%$ of the true value $H$, for the fields considered here; and it was $H'$ which
was monitored to check that the soliton had the required Hopf number.

The initial configurations were taken to be axially-symmetric, with the symmetry
axis chosen so as not to be aligned with the lattice.  Axisymmetric fields \cite{GH97}
are labelled by two integers $(m,n)$, with $mn=H$; for each value of $H$,
various possibilities for $(m,n)$ were tried, and the one
leading to the lowest-energy soliton selected. For example, in the $H=4$ case,
the lowest-energy configuration is of type $(2,2)$, as in the continuum case
\cite{HS00}, and it retains the symmetry of the initial configuration.
For $H=6$, by contrast, the symmetry of the initial axisymmetric $(3,2)$
configuration is lost. In some cases (such as $H=2$), the initially non-aligned
soliton rotates during the minimization so as to become aligned with the lattice.
In other cases (such as $H=1$), the non-aligned soliton has a lower energy
than the corresponding aligned one (which one can obtain by starting with an
aligned initial configuration).  The picture seems to be that there are even more
local minima, in each topological class, than in the continuum case ---
with these additional solutions arising because of the anisotropy of the lattice.

The results of the numerical experiments may be summarized as follows. For each
value of the Hopf number $1\leq H\leq6$, there exists at least one stable local
minimum of $E[\phi]$, and the minimum energies are plotted in Figure 1, on a
log-log scale. The dashed line is the line through the $H=1$ data point having
slope $3/4$; so it is clear that the same power law $E\approx c H^{3/4}$
holds as in the continuum Skyrme-Faddeev system.
\begin{figure}[bth]
\begin{center}
\includegraphics[scale=0.7]{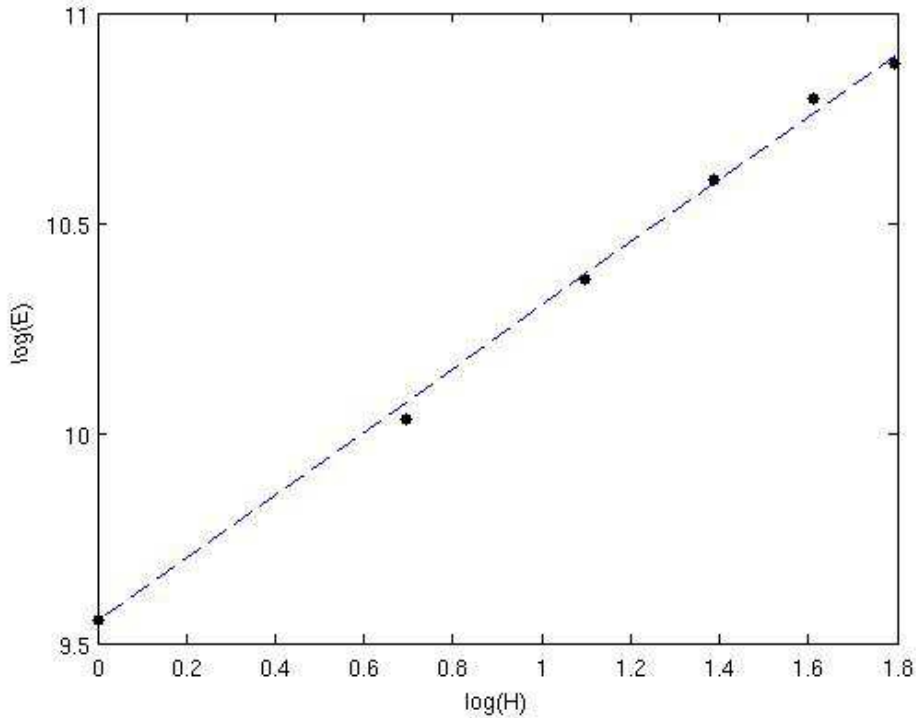}  
\end{center}
\caption{The energy $E$ of minimal-energy configurations, as a function
of the Hopf number $H$, for $1\leq H\leq6$. The dashed line has slope 3/4.}
\label{Fig1}
\end{figure}
The corresponding configurations $\phi$ are illustrated in Figure 2. Recall that
the boundary condition is $\vec\phi\to(0,0,1)$ as $r\to\infty$, and in the
continuum we can visualize the soliton as being located on the closed curve
$\phi_3=-1$ corresponding to the antipodal point on the target space $S^2$.
On the lattice, we obtain an analogous picture by plotting the surfaces
$\phi_3=k$ for some suitable constant $k$.  These surfaces are plotted
in Figure 2, using $k=-0.6$.  In these pictures, the lattice spacing is unity;
so one sees that the solitons are typically spread over a width of 10--20 lattice
units (depending on their shape and on the value of $H$).  It is remarkable
that the pictures closely resemble the corresponding ones in the continuum case
\cite{BS99, HS00}, despite the systems being quite different.
\begin{figure}[thb]
\begin{center}
\includegraphics[scale=0.8]{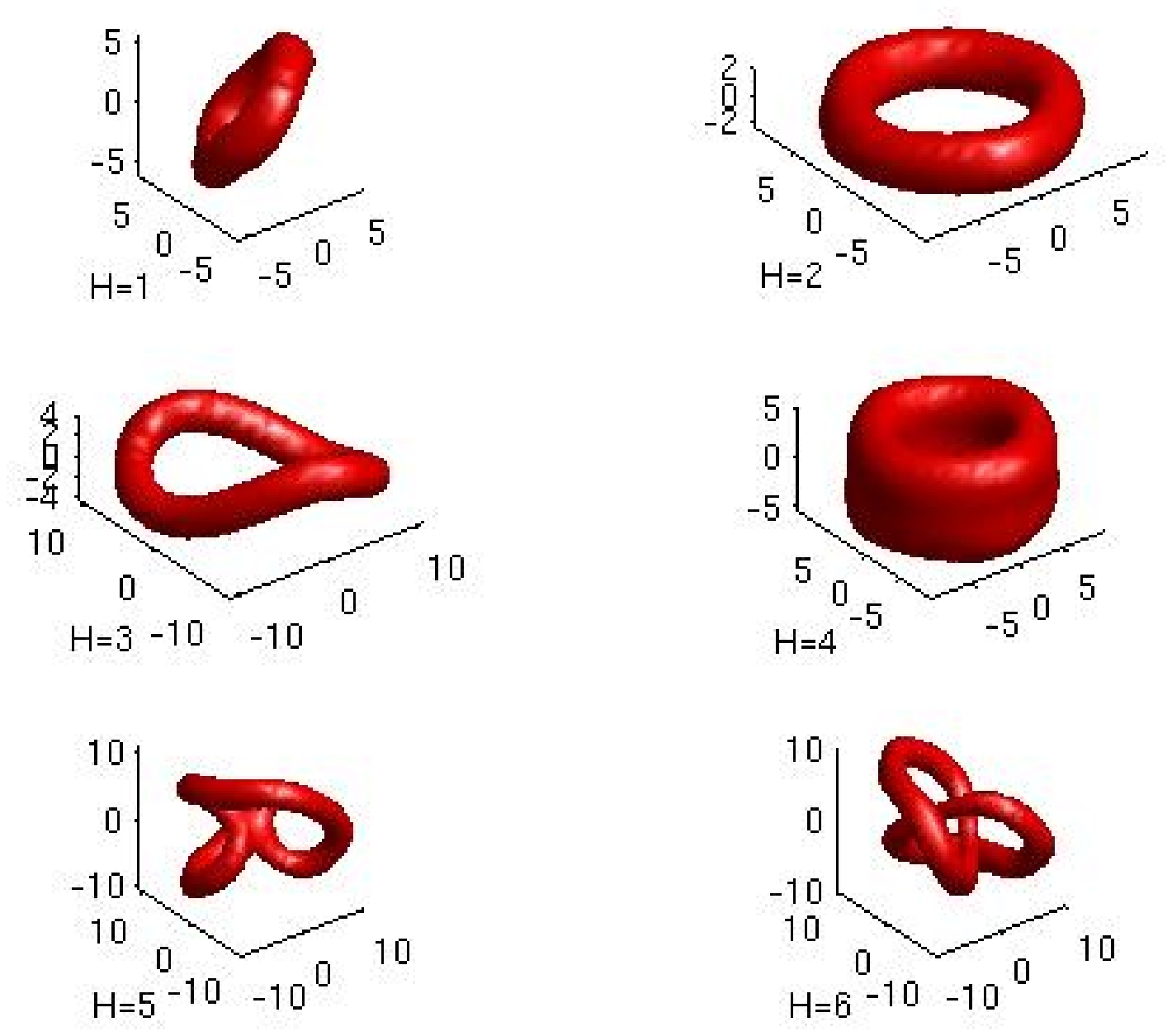}  
\end{center}
\caption{The surfaces $\phi_3=-0.6$ for minimal-energy configurations,
         $1\leq H\leq6$.}
\label{Fig2}
\end{figure}

These results are just for one particular choice of the inter-site potential,
namely (\ref{fxi}).  But various other choices of $f$ have been investigated
as well, and they lead to similar results: for example, the choice \cite{W95}
\begin{equation} \label{f-log}
   f(\xi) = -\log\xi = \sum_{n=1}^{\infty} \frac{1}{n} (1-\xi)^n\,,
\end{equation}
which has the feature that $E[\phi]\to\infty$ as $\phi$ approaches any
discontinuous configuration. The conclusion, therefore, seems to be that
solitons in 3-dimensional O(3) models have ``universal'' features:
provided they exist at all,
their appearance, as well as the 3/4 power law for their energy, are the same
for a wide variety of choices of dynamics.  It might be interesting to
investigate whether there are analytic reasons for generic features such
as these.


\end{document}